# Reputation, Risk, and Trust on User Adoption of Internet Search Engines: The Case of DuckDuckGo

Antonios Saravanos[1(✉) \[0000-0002-6745-810X\]], Stavros Zervoudakis[1], Dongnanzi Zheng[1], Amarpreet Nanda[1], Georgios Shaheen[1], Charles Hornat[1], Jeremiah Konde Chaettle[1], Alassane Yoda[1], Hyeree Park[1], Will Ang[1]

[1] New York University, New York, NY, USA
{saravanos, zervoudakis, dz40, an83, gs3777, cfh1, jk7231, ay2193, hp2240, xa250}@nyu.edu

**Abstract.** This paper investigates the determinants of end-user adoption of the DuckDuckGo search engine coupling the standard UTAUT model with factors to reflect reputation, risk, and trust. An experimental approach was taken to validate our model, where participants were exposed to the DuckDuckGo product using a vignette. Subsequently, answering questions on their perception of the technology. The data was analyzed using the partial least squares-structural equation modeling (PLS-SEM) approach. From the nine distinct factors studied, we found that 'Performance Expectancy' played the greatest role in user decisions on adoption, followed by 'Firm Reputation', 'Initial Trust in Technology', 'Social Influence', and an individual's 'Disposition to Trust'. We conclude by exploring how these findings can explain DuckDuckGo's rising prominence as a search engine.

**Keywords:** end-user adoption, DuckDuckGo, privacy-conscious search engine, trust, risk.

## 1 Introduction

In this paper, we seek to identify the determinants of end-user adoption of the privacy-conscious search engine DuckDuckGo – for those who "are put off by the thought of their every query being tracked and logged" – where there is "absolutely zero user tracking" [5]. We saw the emergence of DuckDuckGo in 2008, recognized as "the first privacy-focused search engine" [9]. The product is designed to cater to a growing number of technology users who value their privacy. The popularity of DuckDuckGo is evident from simply looking at its usage statistics. The company has experienced remarkable growth, going from an annual total of 16,413,461 search queries in 2010, to an annual total of 35,304,278,270 search queries in 2021 [4]. The solution serves as an alternative to the traditional search engines, such as Google, Yahoo, and Bing. While, to the naïve, these search engines may appear to be free, they contain within them a hidden cost: the personal information one imparts to these companies. Certainly, user skepticism regarding the gathering, retaining, and sharing of information by organizations such as Bing and Google "may lead searchers to seek other search engines as alternatives" [3]. Indeed, "just as a car buyer might choose a Volvo over a Ford because the Volvo is said to have better crash impact protection than the Ford, so too might a search engine user choose DuckDuckGo over Google because of the privacy DuckDuckGo offers" [12]. Increasingly we find that there is a newfound awareness amongst users with respect to the tradeoff introduced by search engines: "users are waking up, and search privacy is making its way to the mainstream" [9]. Given DuckDuckGo's rising standing and widespread adoption, there is value in identifying the main determinants of user behavioral intention as well as identifying their respective magnitude.

## 2 Materials and Methods

In this section, we describe the development of the model and hypotheses that were used to investigate user adoption of the DuckDuckGo search engine. We then go on to outline the experimental approach that was taken to evaluate that model, present the data collection process and, lastly, describe the sample demographics.

### 2.1 Model and Hypothesis Development

Correspondingly, a model was developed for the specific technology we are evaluating – a privacy-conscious search engine – which can be seen in Fig. 1. Our model was based on the work of Venkatesh et al. [16] and their Unified Theory of Acceptance and Use of Technology (hereafter UTAUT), which is one of the contemporary models designed to provide insight into user technology adoption decisions. Correspondingly, we generate the following hypotheses:

> H1: 'Performance Expectancy' positively influences 'Behavioral Intention'.
> H2: 'Effort Expectancy' positively influences 'Behavioral Intention'.
> H3: 'Social Influence' positively influences 'Behavioral Intention'.
> H4: 'Facilitating Conditions' positively influence 'Behavioral Intention'.

To this foundation we seek to incorporate into our model the concept of risk. We look to the work of Miltgen et al. [10] and, accordingly, incorporate their construct of "Perceived Risks", concurrently proposing the following hypothesis:

> H5: 'Perceived Risks' positively influence 'Behavioral Intention'.

We also seek to incorporate trust and, to that end, look to Kim et al.'s [8] initial trust model, where the authors propose a construct to reflect initial consumer trust in a technology as well as the antecedents of that, which include a firm's reputation and an individual consumer's personal propensity to trust. As a strategy of how to integrate this construct into our model, we take the approach of Oliveira et al. [11] and subsequently offer the following hypotheses:

> H6: 'Initial Trust' positively influences 'Behavioral Intention'.

H7: 'Firm Reputation' positively influences 'Initial Trust'.
H8: 'Firm Reputation' positively influences 'Behavioral Intention'.

Next, we seek to connect trust and risk into our model and look to the work of Miltgen et al. [10] as precedence, which also links trust to an individual's perceived ease of use and usefulness of the technology. Fittingly, proposing the following hypotheses:

H9: 'Initial Trust' positively influences 'Perceived Risks'.
H10: 'Initial Trust' positively influences 'Performance Expectancy'.
H11: 'Initial Trust' positively influences 'Effort Expectancy'.

Finally, we seek to incorporate how trust of the government may influence the perception of risk, as described by Bélanger and Carter [2]. Accordingly, we propose the following hypotheses:

H12: 'Trust of the Government' positively influences 'Perceived Risks'.
H13: 'Disposition to Trust' positively influences 'Trust of the Government'.
H14: **'**Disposition to Trust' positively influences 'Initial Trust'.

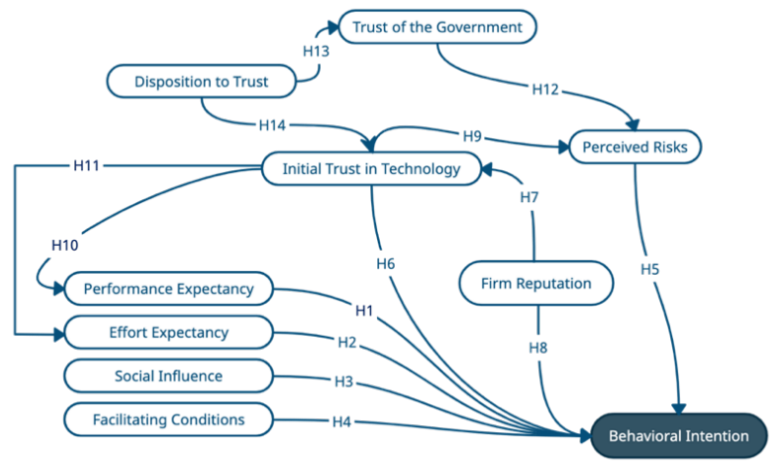

**Fig. 1.** Illustration of our proposed theoretical framework based on the work of Venkatesh et al. [16], Lancelot-Miltgen et al. [10], Kim et al. [8], and Bélanger and Carter [2].

### 2.2 Data Collection and Sample Demographics

An experimental approach was taken. Correspondingly, a questionnaire was developed based on the items provided by the respective authors of the respective constructs to measure user perception of the DuckDuckGo search engine. The questionnaire also included questions to capture participant demographics and ascertain both prior experience using search technology and participant attention to the experiment. Following the obtaining of informed consent, participants were presented with a vignette and animated gif illustrating DuckDuckGo. Subsequently, participants were asked to complete the above-mentioned questionnaire. A total of 322 participants were solicited using Amazon Mechanical Turk. Of those, 81 were removed as they failed the attention checks; this left a total of 241 participations relevant to this study (following the approach of Saravanos et al. [14]). The participant characteristics are outlined in Table 1.

**Table 1.** Participant Demographics.

| Characteristic | Category | N | Percentage |
| --- | --- | --- | --- |
| Age | 18-25 | 9 | 3.73% |
| | 26-30 | 31 | 12.86% |
| | 31-35 | 40 | 16.60% |
| | 36-45 | 68 | 28.22% |
| | 46-55 | 46 | 19.09% |
| | 56 or older | 44 | 18.26% |
| | Prefer not to answer | 3 | 1.24% |
| Gender | Female | 100 | 41.49% |
| | Male | 134 | 55.60% |
| | Other | 2 | 0.83% |
| | Prefer not to answer | 5 | 2.07% |

## 3   Analysis and Results

To analyze the collected data, we followed the technique prescribed by Hair et al. [4]; specifically, we used PLS-SEM coupled with the SmartPLS3.3.2 [13] software. Hair et al. [7] write that "PLS-SEM assessment typically follows a two-step process that involves separate assessments of the measurement models and the structural model". Initially, one "measures' reliability and validity according to certain criteria associated with formative and reflective measurement model specification" [7]. This involved the assessment of convergent validity, construct reliability, and discriminant validity. The first of these, convergent validity, saw us examine the factor loadings followed by the average variance extracted (AVE) and the removal of any manifest variables that had values that were lower than 0.7 with respect to both of these. Following the removal of those items, those remaining were statically significant with a p-value of less than 0.05 after bootstrapping with 7000 subsamples. Construct validity was established by ensuring that both composite reliability (CR) and Cronbach's Alpha were above 0.7. Satisfactory discriminant validity was found through the use of cross-loadings and the Fornell-Larcker criterion.

Subsequently, we examined the structural model (see Table 2). The respective $R^2$ values (see Table 2) show that our model explains: 'Behavioral Intention', 'Effort Expectancy', 'Initial Trust', 'Perceived Risks', 'Performance Expectancy', and 'Trust of the Government' (per the criteria noted by Falk and Miller [6]). We find that the 'Performance Expectancy' (β=0.4302; p<0.01) and 'Social Influence' (β = 0.1345; p<0.05) constructs were statistically significant and played the greatest and fourth-greatest roles in determining user adoption respectively. In other words, the quality of the search results (i.e., 'Performance Expectancy') was the primary determinant of user adoption, and 'peer pressure' (i.e., 'Social Influence') was the fourth greatest. Interestingly, the 'Effort Expectancy' and 'Facilitating Conditions' constructs were not statistically significant. Furthermore, the results revealed that DuckDuckGo's reputation (β=0.4134; p<0.01) was the second most important factor (i.e., 'Firm Reputation') with respect to user adoption decisions and, in relative magnitude, almost equal to 'Performance Expectancy'. In other words, DuckDuckGo's perceived reputation was almost as important to users as the quality of the search results yielded using this tool. Trust in the technology, reflected through the 'Initial Trust' (β=0.2580; p<0.01) construct, played the third-greatest role, and 'Disposition to Trust' (β=0.0279; p<0.05) the fifth-greatest role.

**Table 2.** Results for the Structural Model.

| Path | (Direct) β | (Total) β | (Direct) t-Value | (Total) t-Value |
|---|---|---|---|---|
| *Behavioral Intention ($R^2$=55.05%)* | | | | |
| Disposition to Trust | - | 0.0279* | - | 2.1081 |
| Effort Expectancy | 0.0033 | 0.0033 | 0.0554 | 0.0554 |
| Facilitating Conditions | 0.0397 | 0.0397 | 0.6634 | 0.6634 |
| Firm Reputation | 0.2146* | 0.4134** | 2.2019 | 6.6875 |
| Initial Trust | 0.0801 | 0.2580** | 1.0020 | 2.8875 |
| Performance Expectancy | 0.4302** | 0.4302** | 6.9456 | 6.9456 |
| Perceived Risks | - 0.0151 | - 0.0151 | 0.2479 | 0.2479 |
| Social Influence | 0.1345* | 0.1345* | 2.1680 | 2.1680 |
| Trust of the Government | - | 0.0007 | - | 0.1732 |
| *Effort Expectancy ($R^2$=18.75%)* | | | | |
| Disposition to Trust | - | 0.0462* | - | 2.4671 |
| Firm Reputation | - | 0.3337** | - | 5.9424 |
| Initial Trust | 0.4331** | 0.4331** | 6.8418 | 6.8418 |
| *Initial Trust ($R^2$=65.07%)* | | | | |
| Disposition to Trust | 0.1068* | 0.1068* | 2.5109 | 2.5109 |
| Firm Reputation | 0.7705** | 0.7705** | 24.6209 | 24.6209 |
| *Perceived Risks ($R^2$=44.80%)* | | | | |
| Disposition to Trust | - | - 0.0948** | - | 2.6064 |
| Firm Reputation | - | - 0.5030** | - | 9.3417 |
| Initial Trust | - 0.6529** | - 0.6529** | 11.7024 | 11.7024 |
| Trust of the Government | - 0.0492 | - 0.0492 | 0.9938 | 0.9938 |
| *Performance Expectancy ($R^2$=15.00%)* | | | | |
| Disposition to Trust | - | 0.0414* | - | 2.2815 |
| Firm Reputation | - | 0.2984** | - | 5.5853 |
| Initial Trust | 0.3873** | 0.3873** | 6.2601 | 6.2601 |
| *Trust of the Government ($R^2$=25.89%)* | | | | |
| Disposition to Trust | 0.5088** | 0.5088** | 9.7178 | 9.7178 |

\* $p<0.05$; \*\* $p < 0.01$.

## 4      Discussion and Conclusions

In this study we hypothesized that nine distinct factors would impact user behavioral intention for privacy-focused search engine technology adoption (see Table 3). Of these, four factors were found to have zero impact with respect to user adoption. The first two – the amount of effort needed to use the technology (i.e., 'Effort Expectancy') and the availability of (technical) support (i.e., 'Facilitating Conditions') – were not surprising findings. Certainly, we can attribute this result to the technology we study being very simple to use, and to the fact that all major search engines offer a similar interface. Given that almost all participants (98.3%) reported that they used search engines daily, it is reasonable to conclude that they would perceive the use of DuckDuckGo as effortless and one that they would not require third-party support. What was surprising was that their perception of risk (i.e., 'Perceived Risks') with respect to the technology did not play a role in user decisions to adopt. We attribute the lack of significance in the strength played by the firm's reputation (which was found to be the second most significant factor in determining user adoption). Lastly, we look at individual trust in the government (i.e., 'Trust of the Government'). This can be perhaps explained by the (United States) government having no (substantial) history of monitoring individual search activities. Therefore, this makes it not a significant concern on the part of users.

Five factors were found to have a statistically significant effect on user behavioral intention with respect to adoption: 'Performance Expectancy' was found to have the greatest effect on consumer behavioral intention, followed by 'Firm Reputation', 'Initial Trust', 'Social Influence', and 'Disposition to Trust'. The finding that the perceived usefulness (i.e., 'Performance Expectancy') played the greatest role in user decisions in and of itself is not surprising, given that adoption studies frequently see this factor playing a significant role. Thus, the quality of the search results are what users look to first with respect to adoption. In other words, if DuckDuckGo is to compete with the major search engines, such as Google, Bing, and Yahoo, it must ensure a comparable quality in the results it returns.

The next factor was DuckDuckGo's reputation (i.e., 'Firm Reputation'), indicating that, with such a product, users look to the brand when deciding whether to use, and then subsequently assess their initial trust in the technology (i.e., 'Initial Trust'). Hence, it is crucial for DuckDuckGo to maintain the quality of its brand and trust in the technology that they offer if they want to preserve and expand their user base. This also opens the door to future research, which may seek to identify the tradeoff between the quality of the result and the firm's reputation and trust in the product.

Next on the list was the role that the opinion of their peers (i.e., 'Social Influence') plays in user decisions on whether to adopt. Accordingly, DuckDuckGo should seek to promote its technology through existing users (e.g., peer pressure) and marketing. The final factor examined was an individual's unique disposition to trust (i.e., 'Disposition to Trust'). Thus, whether a user is, by their nature, trusting would influence whether they use such a product, with those that are not trusting being more inclined to adopt. Consequently, DuckDuckGo may want to focus its efforts on such individuals.

In conclusion, our findings explain the rising usage of DuckDuckGo; while the quality of search results is a strong factor in determining adoption, DuckDuckGo's reputation, the trust placed in the technology, and an individual's disposition to trust, also play a prominent role in users' adoption decisions.

**Table 3**. Results of Hypothesis Testing.

| Hypothesis | Causal Path | Remarks |
| --- | --- | --- |
| H1 | Performance expectancy → Behavioral intention | Supported |
| H2 | Effort expectancy → Behavioral intention | Not Supported |
| H3 | Social influence → Behavioral intention | Supported |
| H4 | Facilitating conditions → Behavioral intention | Not Supported |
| H5 | Perceived risks → Behavioral intention | Supported[1] |
| H6 | Initial trust → Behavioral intention | Supported[1] |
| H7 | Firm reputation → Initial trust | Supported |
| H8 | Firm reputation → Behavioral intention | Supported |
| H9 | Initial trust → Perceived risks | Supported |
| H10 | Initial trust → Performance expectancy | Supported |
| H11 | Initial trust → Effort expectancy | Supported |
| H12 | Trust of the government → Perceived risks | Not Supported |
| H13 | Disposition to trust → Trust of the government | Supported |
| H14 | Disposition to trust → Initial trust | Supported |

[1] Supported via total effect.

## 4.1 Limitations and Future Research Directions

With respect to this work, we note three limitations that should be highlighted. Tangentially, we present how these limitations also offer direction for future research on the topic. The first limitation relates to the effect that participant culture can have on consumer intention to accept a technology. Clearly, there is evidence (e.g., [1, 15, 17]) that raises this as an area of concern. In this paper, we restricted our sample solely to participants from the United States. Accordingly, the study of other cultures may lead to different findings and broaden our research. The second limitation refers to the method through which we exposed participants to the technology. Rather than having them interact with the DuckDuckGo search engine, they were offered a vignette and animated gif that highlighted the characteristics of the solution. It is possible that actual interaction with the technology could lead to different user perceptions and that, over time, those perceptions could change. The third limitation is with respect to our research focusing on users' intention to adopt, and accordingly did not investigate their actual usage. These limitations also identify how this work can be further developed: looking at users from different cultures; having participants actually interact with the DuckDuckGo product; and investigating how that interaction actually transforms into usage.

**Acknowledgments.** This research was funded in part through the New York University School of Professional Studies Full-Time Faculty Professional Development Fund.

**References**

1. Bandyopadhyay, K., Fraccastoro, K.: The effect of culture on user acceptance of information technology. Communications of the Association for Information Systems. 19, 522–543 (2007). https://doi.org/10.17705/1CAIS.01923.
2. Bélanger, F., Carter, L.: Trust and risk in e-government adoption. The Journal of Strategic Information Systems. 17, 2, 165–176 (2008). https://doi.org/10.1016/j.jsis.2007.12.002.
3. Burnett, A.: A Baker's Dozen of Tips for Better Web Searches. Digital Commons @ University of Georgia School of Law (2020), https://digitalcommons.law.uga.edu/cgi/viewcontent.cgi?article=1077&context=cle, last accessed 2022/04/05.
4. DuckDuckGo: DuckDuckGo Traffic, https://duckduckgo.com/traffic, last accessed 2022/05/25.
5. Duò, M.: 21 Alternative Search Engines to Use in 2022, https://kinsta.com/blog/alternative-search-engines/, last accessed 2022/04/05.
6. Falk, R.F., Miller, N.B.: A Primer for Soft Modeling. University of Akron Press, Ohio (1992).
7. Hair, J.F. et al.: PLS-SEM: Indeed a silver bullet. Journal of Marketing Theory and Practice. 19, 2, 139–152 (2011). https://doi.org/10.2753/MTP1069-6679190202.
8. Kim, G. et al.: Understanding dynamics between initial trust and usage intentions of mobile banking. Information Systems Journal. 19, 3, 283–311 (2009).
9. Kite-Powell, J.: This New Privacy-First Search Engine Keeps Your Searches Private, https://www.forbes.com/sites/jenniferhicks/2020/02/12/this-new-privacy-first-search-engine-keeps-your-searches-private/, last accessed 2022/01/28.
10. Lancelot Miltgen, C. et al.: Determinants of end-user acceptance of biometrics: Integrating the "Big 3" of technology acceptance with privacy context. Decision Support Systems. 56, 103–114 (2013). https://doi.org/10.1016/j.dss.2013.05.010.
11. Oliveira, T. et al.: Extending the understanding of mobile banking adoption: When UTAUT meets TTF and ITM. International Journal of Information Management. 34, 5, 689–703 (2014). https://doi.org/10.1016/j.ijinfomgt.2014.06.004.
12. Pasquale, F.: Privacy, antitrust, and power. George Mason Law Review. 20, 4, 1009–1024 (2012).
13. Ringle, C.M. et al.: SmartPLS 3. Bönningstedt: SmartPLS. (2015). https://www.smartpls.com/, last accessed 2022/05/22.
14. Saravanos, A. et al.: The hidden cost of using Amazon Mechanical Turk for research. In: Stephanidis, C. et al. (eds.) HCI International 2021 - Late Breaking Papers: Design and User Experience. pp. 147–164 Springer International Publishing, Cham (2021). https://doi.org/10.1007/978-3-030-90238-4_12.
15. Srite, M., Karahanna, E.: The role of espoused national cultural values in technology acceptance. MIS Quarterly. 30, 3, 679–704 (2006). https://doi.org/10.2307/25148745.
16. Venkatesh, V. et al.: User acceptance of information technology: Toward a unified view. MIS Quarterly. 27, 3, 425–478 (2003). https://doi.org/10.2307/30036540.
17. Yuen, Y.: Internet banking acceptance in the United States and Malaysia: A cross-cultural examination. Marketing Intelligence & Planning. 33, 3, 292–308 (2015). https://doi.org/10.1108/MIP-08-2013-0126.